\documentclass{article}

\usepackage[preprint]{neurips_2020}
\usepackage{natbib}
\setcitestyle{numbers}
\usepackage[utf8]{inputenc} 
\usepackage[T1]{fontenc}    
\usepackage{hyperref}       
\usepackage{url}            
\usepackage{booktabs}       
\usepackage{amsfonts}       
\usepackage{nicefrac}       
\usepackage{microtype}      

\title{AI in the “Real World”: Examining the Impact of AI Deployment in Low-Resource Contexts}

\author{%
  Chinasa T. Okolo\thanks{cs.cornell.edu/~chinasa} \\
  Department of Computer Science\\
  Cornell University\\
  Ithaca, NY 14853 \\
  \texttt{chinasa@cs.cornell.edu}
}

\begin{document}

\maketitle

\begin{abstract}
  As AI becomes integrated throughout the world, its potential for impact within low-resource regions around the Global South have grown. AI research labs from tech giants like Microsoft, Google, and IBM have a significant presence in countries such as India, Ghana, and South Africa. The work done by these labs is often motivated by the potential impact it could have on local populations, but the deployment of these tools has not always gone smoothly. This paper presents a case study examining the deployment of AI by large industry labs situated in low-resource contexts, highlights factors impacting unanticipated deployments, and reflects on the state of AI deployment within the Global South, providing suggestions that embrace inclusive design methodologies within AI development that prioritize the needs of marginalized communities and elevate their status not just as beneficiaries of AI systems but as primary stakeholders.
\end{abstract}

\section{Introduction}
While the surge of AI being deployed for use in low-resource contexts is promising, there are often things overlooked throughout the development and implementation stages of these projects. In this paper, we discuss relevant risks and concerns of AI research in the Global South, highlight a short case study involving a project that had unanticipated impacts of an AI system, and propose methods to increase the use of participatory methods in the creation of AI systems and decrease the risks of harming local communities through these technologies.

\section{Background}

\subsection{AI for the Global South}
As opportunities for artificial intelligence (AI) increase due to advances in computing power and the expansion of data collection practices, its potential to transform life for marginalized populations within low-resource contexts has been recognized. So far, AI has been applied to a variety of contexts in agriculture, education, finance, healthcare, and manufacturing. More specifically, within AI research focusing on contexts for the Global South, we have seen projects focused on optimizing machine learning for smaller devices \citep{gopinath2019compiling}, developing data-driven agriculture techniques to improve harvest yields and prevent waste \citep{vasisht2017farmbeats}, automating labeling of pathology reports \citep{saib2020hierarchical}, and using social media to analyze the mental health of users \citep{sharma2020engagement}.

\subsection{AI Research Labs in the Global South}
With AI becoming an aspect of daily life for people around the world, large tech companies have noticed the opportunities arising from AI and its respective applications and have begun to expand their research labs into countries within the Global South. IBM, perhaps one of the earliest tech companies to establish research labs, has a significant presence within the Global South. IBM Research has labs across the world in Africa: Nairobi, Kenya, Johannesburg, South Africa; South America: Rio de Janiero and Sao Paulo, Brazil; and in Southeast Asia: Bangalore and Delhi, India \citep{ibmresearchlabs}. On the other hand, both Microsoft and Google have significantly fewer research labs in the Global South with locations in Accra, Ghana (Google), Nairobi, Kenya (Microsoft), and Bangalore, India (Google, Microsoft) \citep{googleindia, cisse2018look, msrafrica}. Other notable tech companies such as Facebook and Amazon operate data centers and/or offices in countries within the Global South, but have no established research labs in these locations.

\subsection{Participatory AI}
In many cases, the success of an AI system is dependent on the environment it is integrated into. To improve the likelihood of success, participatory design methods taken from the field of human-computer interaction \citep{muller1993participatory} have become more popular in the development of AI systems. Participatory design encourages all stakeholders to be involved in collaborating with developers and researchers to inform and guide the development of solutions and is not only limited to technical systems \citep{cornwall1995participatory}. However, Brattieg et. al highlight the challenges AI poses to participatory design methods, especially in the understanding of the respective capabilities and limitations of AI as a technique \citep{bratteteig2018does}. Not only is it necessary to consider the infrastructural differences in the capacity of an environment to handle AI, but factors such as literacy, culture, values, norms, and community needs are crucial in designing adaptable AI tools. As AI systems transition out of well-funded and high-resourced research labs into contrasting environments, maintaining the philosophies of participatory design will hopefully aid in their success.

\section{Case Study}
With the expansion of premier AI research labs into countries throughout the Global South, the development of AI applications to solve local issues in healthcare have become more common. Motivated by issues with screening for eye disease in Southeast Asia, researchers at Google AI in the United States applied deep learning techniques in the creation of an algorithm to detect diabetic retinopathy from retinal scans from patients in the United States and India \citep{gulshan2019performance}. To evaluate this algorithm in clinical settings, researchers from Google Health conducted a human-centered observational study in partnership with clinics in Thailand to analyze how the introduction of this algorithm impacted clinical workflows and the factors that affected algorithmic performance of the AI system \citep{beede2020human}. 

Despite the researchers performing fieldwork pre- and post-deployment of the algorithm, there were significant issues that occurred in the integration of the system. A major finding from this study showed that the system could not handle low-quality images taken in the Thai clinics, as the underlying algorithm had originally been trained on high-quality lab images. The inability of the system to handle images produced from the clinic led to major disruptions both during and after the screening process. The clinic workers who were already overburdened by a large patient load had to take more photos to produce a suitable image for the AI system, lengthening the time spent per visit. Additionally, the prospect of receiving a false positive diagnosis could cause both emotional and financial harm to patients who had to deal with the stress of a "positive" diagnosis and take more time off of work to pursue follow up appointments at a clinic further from their homes. Additional factors such as low internet speed/connectivity, lack of screening rooms that could take high-quality images, and the availability of medical professionals suited to analyze retinal images also impacted the success of this AI system.

When reflecting on this study, it becomes clear that many of the human-centered approaches referenced in the paper would have benefited the development of the system had they occurred pre-deployment. The lack of understanding around the existing workflows of analyzing diabetic retinopathy in the participating clinics led to the failures of this AI system once deployed. Both the patients and and the nurses in these clinics provide valuable contributions that would've prevented the oversights seen in this study. However, the benefit of this study lies in its contribution as the first known study to use human-centered methodologies to evaluate deep learning systems applied in clinical contexts. As similar research evolves within this field, we hope that these techniques will encourage critical reflection during the AI development process that takes into account the implications and underlying harms their systems can cause on vulnerable populations.

\section{Anticipated risks and known harms of AI research}

\subsection{Who is best placed to anticipate and address potential research impacts?}
\label{anticipateimpact}
To begin the process of figuring out the anticipated risks and known harms of AI research, it is important to include the voices of local stakeholders such as rural residents, farmers, community health workers, etc. Even within the research labs situated in the Global South, the researchers themselves are usually not the intended user populations of the technologies they develop, so it becomes even more critical to ensure that local communities are actively involved in the development of these AI systems. The background knowledge these stakeholders bring to projects cannot be understated. Their first-hand experience with systems and processes augmented by AI is important to understand how and why AI could benefit in the first place. Another overlooked community within this topic is non-computer scientists such as UX/UI designers and more notably, anthropologists and sociologists. Many social scientists spend years assimilating into native populations to become familiar with their culture and practices before conducting studies. While it may not be necessary to spend as much time for technical studies, adopting some methodologies from social scientists and utilizing user research tools of HCI researchers may prove helpful in improving current participatory methods used in AI deployments.

\subsection{The role of AI researchers and the AI research community}
While the AI development life cycle includes many stakeholders, AI researchers should not be limited to just the technical development stage. Within their work, we believe that it is important for AI researchers to:

\begin{enumerate}
  \item \emph{Amplify} the voices of communities they are intending to impact by (1) encouraging their inclusion in the design, development, and deployment of AI tools, aligning with participatory design methods and (2) providing resources for communities to be self-reliant in maintaining AI tools. The second action point is much harder to achieve, but significant progress is continuing to be made as outlined in the later half of this section.
  \item \emph{Build}, \emph{develop}, and \emph{deploy} AI tools with intention. A helpful start to this process can begin by integrating theoretical principles of Decolonial AI, as pioneered by Mohamed et. al \citep{mohamed2020decolonial}, into every stage of the AI tool-building and evaluation process. The first tactic of supporting a "critical technical practice" of AI, is especially relevant. Moving from a standpoint of dropping AI tools into environments without considering the underlying implications to a point where these considerations are put at the forefront, will make a significant difference.
  \item  Create computational tools that are \emph{effective in a variety of contexts}. It is the most basic responsibility for AI to work for all people irrespective of ethnic background, socioeconomic standing, and place of living. As highlighted in the case study, we see that moving AI out of the lab is not always a simple process. However, we believe that undergoing critical and "self-reflexive" evaluations in the development and deployment of AI can aid in this process.
\end{enumerate}

The prior list is non-exhaustive, but provides a solid base for the AI community to move forward on. As a whole, the AI community has much more to achieve in becoming more inclusive of researchers from underrepresented populations. Work by top research labs has already done a significant job in increasing the number of AI researchers in the Global South. However, there is still a gap in the training needed to get researchers into these highly-coveted positions. Organizations such as the African Institute for Mathematical Sciences (AIMS), Data Science Nigeria, Masakhane, Deep Learning Indaba, Khipu, Black in AI, Latinx in AI, and more have led grassroots initiatives mentoring and educating underrepresented researchers throughout the Global South.

\section{Collective and individual responsibility in AI research}

\subsection{Risk identification within AI}
The Google AI study highlighted the shortcomings that can occur in the deployment of AI systems despite being situated relatively close to the target user population. Risk identification is not something that can be done retroactively and has to be ingrained within the development team from the start of a project. To begin, conducting extensive literature reviews is a great way to familiarize technical collaborators with participatory design methodologies. As mentioned in Section \ref{anticipateimpact}, the research methodologies of social scientists, along with actively including them in projects will positively aid in the process of risk identification. Conducting ethnographic and feasibility studies before technical deployments is always a great step. Not only will the developers learn more about the needs and values of their target community, but these studies will inform the project team of external factors (sociocultural, infrastructural, political, governmental, etc.) that could hinder a successful deployment.

\subsection{Dimensions of concern within AI}
The dimensions of concern within the work of researchers who develop AI-enabled tools can be summarized into four (4) facets: ethical, cultural, feasibility, explainability. Within the field of AI research, topics of fairness, accountability, and transparency have become prominent conversations with multiple conferences, workshops, and symposiums spinning out to support conversations in this field. With the harm AI systems have enacted through exacerbating racial inequities within fields such as healthcare \citep{obermeyer2019dissecting} and policing \citep{babuta2019data, bellamy2019think, hill2020wrongfully}, it is imperative that AI researchers are aware of the ethical implications associated with the potential increase in workload for some stakeholders, the need for this system, and justifications behind deploying this system within the target locale. Developers embed their own set of values within the systems they build and this transfer into different settings may produce a clash. Understanding factors such as cultural belief of technology and how this impacts acceptance of new technologies is important. While the potential for impact may be high, determining technical feasibility is a must. More simply: your solution can't have impact if it doesn't work. Additionally, AI models are commonly unoptimized for low-resource contexts and understanding the infrastructural capacity of a specific institution is helpful in determining feasibility. Explaining AI is often a challenge that impacts both developers and users of AI systems. In contexts where technical knowledge is already low, ensuring that users understand what a system is doing and why is important.

\subsection{Awareness of known harms}
Bringing awareness to known harms caused by related research can primarily be done through conducting literature searches. Unfortunately, within the field of AI, there has been less work focused towards applying theoretical methods, especially within low-resource contexts. This leaves us with a paucity of research demonstrating how AI works when applied in different contexts. Combined with the lack of activity in deploying AI, there is a tendency to only hear about the “successes” or projects from larger, well-funded research institutions. Establishing venues where this type of work can be made more accessible is also a great step towards democratizing the development and utilization of AI by those in low-resource environments. We also believe that improving accountability for AI deployments within the Global South is another way to increase the awareness of known harms by encouraging publications of findings, whether good or bad.

\subsection{Responsiveness to the needs and concerns of affected communities}
As the field of AI expands and deepens its roots into technological ecosystems throughout the Global South, the lack of existing regulation governing the use of AI could potentially open the door to algorithmic abuse of vulnerable populations. Fortunately, researchers can provide the formal understanding of the implications and opportunities associated with AI needed to drive policy and legislation regulating its use. Prior work building frameworks to guide AI deployment within low-resource environments and gauge user perception of AI technologies show promise for informing the design of such AI policies \citep{2020acips}. We believe that integrating the philosophies provided by participatory design and Decolonial AI into the development of AI systems will ensure that the field is responsive to both the needs and concerns of populations affected by related research, enabling marginalized communities to be fully present and active contributors within all aspects of the AI development process.

\section{Conclusion}
As developers continue to build AI technologies with the intent to integrate them into the "real-world", it is important that they prioritize the perspective of users, especially those situated within low-resource environments. The burgeoning bodies of literature and venues dedicated to analyzing the implications of AI research in various contexts has ignited these conversations. Additionally, actions taken by conferences such as NeurIPS to include broader impact statements are a good start in encouraging critical reflection in the research lifecycle, but there is still need for a field-wide paradigm shift among the designers and developers of AI systems.

\bibliographystyle{ACM-Reference-Format}
\bibliography{references}

\end{document}